\documentclass[prb, twocolumn, showpacs, preprintnumbers, amsmath, amssymb, superscriptaddress, aps]{revtex4}

\usepackage{color}
\usepackage{bm}
\usepackage{hyperref}
\usepackage{graphicx}
\usepackage{braket}
\usepackage{MnSymbol}
\usepackage{amssymb}
\hypersetup{colorlinks=true}
\usepackage{dcolumn}
\usepackage{amsmath}
\usepackage{amsfonts}
\usepackage{bm}
\usepackage{color}
\usepackage[normalem]{ulem}

\usepackage{hyperref}
\hypersetup{colorlinks,linkcolor=blue,urlcolor=blue,citecolor=blue}

\newcommand{\p}{\partial}
\newcommand{\s}{\sigma}

\newcommand{\la}{\langle}
\newcommand{\ra}{\rangle}
\newcommand{\rd}{\mbox{d}}
\newcommand{\ri}{\mbox{i}}

\renewcommand{\vec}[1]{{\bm #1}}

\renewcommand{\v}[1]{\textbf{#1}}

\begin{document}
\title{Spin magnetometry as a probe of stripe superconductivity in twisted bilayer graphene}
\author{E.~J.~K\"onig}
\affiliation{Department of Physics and Astronomy, Center for Materials Theory, Rutgers University, Piscataway, NJ 08854, USA}
\author{Piers Coleman}
\affiliation{Department of Physics and Astronomy, Center for Materials Theory, Rutgers University, Piscataway, NJ 08854, USA}
\affiliation{Department of Physics, Royal Holloway, University of London, Egham, Surrey TW20 0EX, UK}
\author{A. M. Tsvelik}
\affiliation{Division of Condensed Matter Physics and Materials Science, Brookhaven National Laboratory, Upton, NY 11973-5000, USA}
 \date{\today }

\begin{abstract} 
The discovery  of alternating superconducting and insulating ground-states in
magic angle graphene has suggested an intriguing analogy with 
cuprate high-$T_c$ materials.
Here we argue that the network states of small angle twisted bilayer 
graphene (TBG) afford a further perspective on the cuprates by 
emulating their 
stripe-ordered phases, 
as in La$_{1.875}$Ba$_{0.125}$CuO$_4$.  We show that 
the spin and valley quantum numbers of stripes in TBG
graphene fractionalize, developing  characteristic signatures in
the tunneling density of states and 
the magnetic noise spectrum of 
impurity spins. 
By examining the coupling between the  charge rivers we determine the superconducting transition
temperature. Our study suggests that magic angle graphene can be used 
for a controlled emulation of stripe superconductivity and quantum sensing
experiments of emergent anyonic excitations.
\end{abstract}


\maketitle
\section{Introduction.} 
The recent observation of superconducting 
and correlated insulating states in twisted bilayer graphene (TBG)~\cite{CaoJarillo2018a,CaoJarillo2018b}
have established twisttronics as a new platform for
studying strongly correlated quantum materials. In particular, the
appearance of superconducting domes, with strange metallic behavior in
the normal state, near putative Mott insulators at the magic twist
angle ($\theta = 1.1^\circ$) establishes a striking parallel between 
phase diagrams of TBG and cuprate {high-$T_c$ superconductors, which} has caused a great excitement in the community. 

\begin{figure}
\includegraphics[width=0.48 \textwidth]{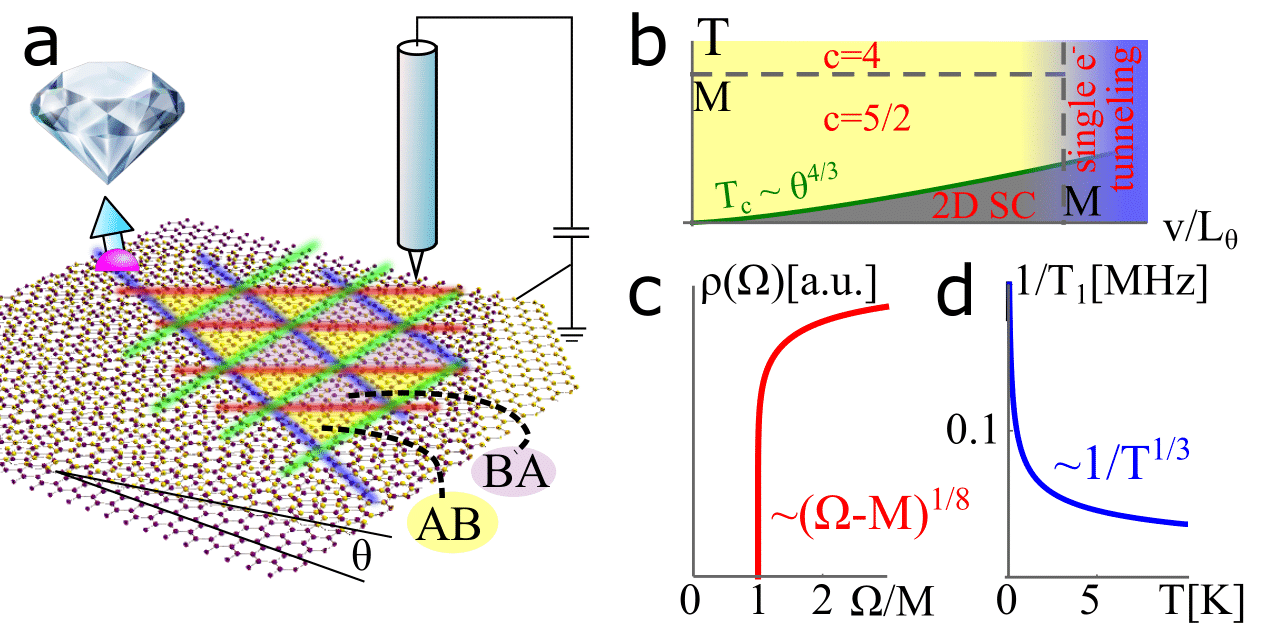}
\caption{{\bf a} Network states in small angle twisted bilayer graphene and illustration of local experimental probes. {\bf b} Phase diagram, in the plane temperature $T$ vs. inverse moir\'e length $1 /L_\theta \propto \sin(\theta/2)$, which contains a fractionalized low energy $U_4(1) \times SU_2(2)$ state with a dynamical gap $M$ in the orbital sector and central charge $c = 5/2$~\cite{xu}. Within this phase, 2D singlet superconductivity (SC) emerges with $T_c$ given by Eq.~\eqref{eq:Tc}. {\bf c} The tunneling density of states displays the dynamical gap $M$ and a characteristic power-law above the gap. {\bf d} Due to the dynamical mass in the orbital sector, impurity spins are subject to robust four-channel Kondo criticality without fine tuning. This results in a characteristic $T^{-1/3}$ divergence of the decoherence rate $1/T_1$ in NV center magnetometry. (For panels {\bf b, c} we used $K = 1$ and for {\bf d} $T_K = 68$K, $r = 5$ nm).} 
\label{fig:summary}
\end{figure}

In this paper we discuss a new analogy between cuprates and TBG,
in the regime of small twist angles, where, in the presence
of a displacement field, theory
predicts~\cite{SanJosePrada2013,mcdonald,WaletGuinea2019,FleischmannShallcross2019,TsimKoshino2020,DeBeuleRecher2020}
that the bulk of the sample has a gap, while gapless excitations
propagate along domain walls separating regions of ``{AB}'' stacking
from regions of ``{BA}'' stacking, Fig.~\ref{fig:summary} {\bf
a}. {The essence of this scenario was recently confirmed by near field spectroscopy~\cite{polaritons}, scanning tunneling microscopy~\cite{HuangLeRoy2018}, and transport~\cite{aharonov,xuGeim2019,YooKim2019} experiments.}
The emergent triangular network of conductive domain walls 
immersed in the insulating bulk is closely analogous to the $x = 1/8$
anomaly of Lanthanum cuprates, particularly
La$_{1.875}$Ba$_{0.125}$CuO$_4$ (LBCO)~\cite{KivelsonRMP}. Current 
theories of stripe order in the cuprates~\cite{BergZhang2007} 
predict that each individual stripe is a Luther-Emery liquid with a spin gap and 
superconducting pairing; global coherence is established by pair
tunneling between the stripes~\cite{qli2007}. 
While this theory  accounts for the gross features of the stripe
phase, it has difficulties  to account for recent high field measurements 
which reveal the development of a low temperature metallic state 
where the standard theory predicts insulating behavior \cite{tranquada,dragana1,dragana2}. 
The discovery  of stripe-like conducting behavior 
in small angle twisted bilayer graphene provides a new platform to 
explore the interplay of stripes with superconductivity. 
Motivated by these observations, here we present a study of many-body
effects in TBG networks.~\cite{xu,ChouNandkishore2019,ChenPereira2020} A single domain wall displays an
unconventional pattern of fractionalization in which the emergent
orbital sector is gapped, but spin and charge excitations remain
critical~\cite{xu} (see Fig.~\ref{fig:summary} {\bf b}
and~\ref{fig:orbital} {\bf d}). We investigate the gapped
tunneling density of states in these wires (Fig.~\ref{fig:summary}
{\bf c}) and propose that the detection of  their fractionalized
spin modes can be achieved by introducing magnetic impurities 
along the domain wall (for example, by
irradiation~\cite{ChenFuhrer2011,JiangAndrei2018}). In particular, 
the pattern of fractionalization induces an overscreened 4-channel Kondo
(4CK) effect which is energetically
protected by the gap in the orbital sector. 
Generally, the signatures of anyons 
at multi-channel Kondo
impurities~\cite{AndreiDestri1984,AffleckLudwig1991} are fragile and
disappear with weak channel asymmetry, but 
in the TBG stripes, these features are protected by the orbital gap,
so this behavior should be robust. We explore the possibility of using
noise magnetometry {and two-dimensional non-linear
spectroscopy~\cite{KuehnHey2011,WernerElsaesser2013,WanArmitage2019,ChoiKim2020}
using} proximitized spin-qubits such as nitrogen-vacancy (NV) centers in
diamond~\cite{TaylorLukin2008}
as a non-invasive local  probe of such multi-channel Kondo
 criticality, see
Fig.~\ref{fig:summary} {\bf d}. These methods are now capable of resolving a 
single electronic spin~\cite{GrinoldsYacoby2013}.  We have also examined 
the  effect of 
coupling of the wires~\cite{xu}
and the emergence of a coherent 2D superconductivity in the TGB
networks, determining the  critical temperature and magnetic field.

This paper contains a section on the physics of a single domain wall, Sec.~\ref{sec:SingleWall}, followed by experimental probes of fractionalization, Sec.~\ref{sec:Exp}, and a discussion of 2D superconductivity, Sec.~\ref{sec:2DSC}. We conclude with a summary and outlook and relegate technical details to five appendices.

\section{A single domain wall}
\label{sec:SingleWall}
\begin{figure}[t]
   \includegraphics[width=0.47 \textwidth]{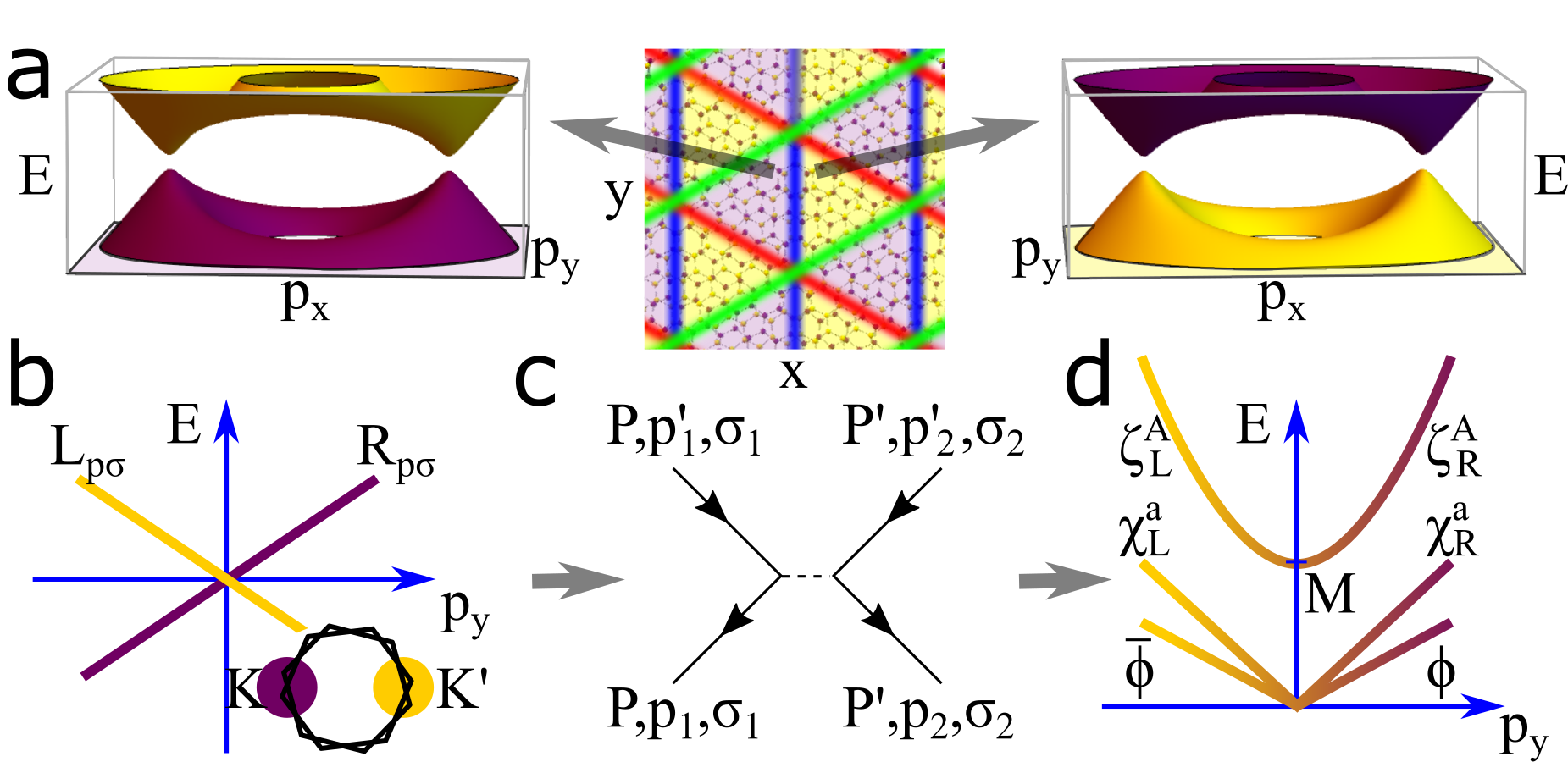}
  \caption{{\bf a} At a domain wall, e.g. at $x = 0$, the Chern number of a given valley (color coded bands) changes by two. Since both topological gaps close at $p_y =0$~\cite{mcdonald}, two degenerate right (left) movers per spin emerge at the $K$-$(K')$ valley, panel {\bf b} and Eq.~\eqref{H0}. Their wavefunction has relative phase $\pi/2$ ($-\pi/2$) between top and bottom layer components. {\bf c} Therefore, the valley quantum number ($P = K, K'$) is conserved at two-body interactions, Eq.~\eqref{currents}, which leads to a dynamical mass generation in the orbital SU$_2$(2) sector. {\bf d} The low-energy theory is thus  U(1)$\times$SU$_2$(2) critical model.}
\label{fig:orbital}
\end{figure}

We adopt the description of the triangular
network~\cite{SanJosePrada2013} in 
TBG {developed by Efimkin and MacDonald}~\cite{mcdonald}.
Each domain wall carries two chiral modes corresponding to the states
centered around the $K$ and $K'$ points of the graphene Brillouin zone,
Fig.~\ref{fig:orbital} {\bf a}. These  chiral fermions carry spin $\s = \pm 1/2$ and {orbital} 
quantum numbers $p= \pm 1$,  so that the Hamiltonian for a single domain wall is 
 \begin{eqnarray}
 H_0 = \int \rd x \Big(-\ri v R^+_{p,\s}\p_x R_{p,\s} + \ri v L^+_{p,\s}\p_x L_{p,\s}\Big). \label{H0}
 \end{eqnarray} 
Here, $v \sim w v_0/u$\ ~\cite{mcdonald} is the velocity of the
modes, where $v_0 \sim 10^6 m/s$ is the bare nodal fermion speed in graphene, $u \sim 0.1 eV$ the interlayer bias~\cite{polaritons} and $w \sim 0.1 eV$~\cite{BistritzerMacDonald2011} the interlayer hopping. The latter scales determine the gap in AB and BA regions and thereby set the UV cut-off for the network model (we refer the reader to Table~\ref{tab:Scales} for a summary of experimental scales). Throughout the manuscript we use natural units $\hbar = 1 = k_B$.
For narrow gate separations, the many-body
physics is determined by a local, screened Coulomb interaction. 
When we project the direct and exchange Coulomb interactions onto the low-lying states of the 1D wires (see Appendix~\ref{app:MicroDerivation} for details), in the experimentally~\cite{polaritons} relevant case of wide domain walls, we obtain the model
 \begin{eqnarray}
 && H_{\rm int} =g \int \rd x  \;2 J_R^0 J_L^0+ \sum_{A ={{x,y}}} J_R^A J_L^A, \nonumber\\
 && J_R^A = \frac{1}{2}R^+_{p,\s}\tau^A_{pq}R_{q,\s}, ~~ J_L^A = \frac{1}{2}L^+_{p,\s}\tau^A_{pq}L_{q,\s}. \label{currents}
 \end{eqnarray}
Here, $\tau^A \in \{ \mathbf 1,\tau_x, \tau_y, \tau_z \}$ represent Pauli matrices in orbital space. Most importantly, only the density and the orbital currents interact.

To a first approximation the domain walls can be treated
as a system of 1D quantum wires, 
neglecting their mutual  interaction. 
The model (\ref{H0},\ref{currents}) is exactly solvable; at low
energies the fixed point behavior becomes orbitally isotropic,
and indistinguishable from the SU(2)
invariant model of Ref.~\onlinecite{xu}. 
The orbital coupling 
is marginally relevant and generates a gap $M$ in the orbital
sector. The gapless sector is described by U(1)$\times$SU$_2$(2)
critical model,~\ref{fig:orbital} {\bf d}. The scaling dimensions of 
the U(1) sector are determined by the Luttinger parameter $K$ which
is, in turn, set by the forward scattering. {The SU$_2$(2) spin physics
is governed by}
\begin{eqnarray}
{ H[SU_2(2)]} &=& {\frac{\pi}{2} \sum_{a = 1}^3 \sum_{\zeta = L/R} \int dx : J^a_{\zeta}(x) J^a_{\zeta}(x):,} \label{eq:SU2}
\end{eqnarray}
where $J^a_{\zeta}(x)$ obey the $su_2(2)$ Kac-Moody algebra and can be
represented by three Majorana fields $J^a_{\zeta}(x) = - i \epsilon
^{abc} \chi_\zeta^b \chi_\zeta^c/2$. 
Physical electrons  
are gapped and largely incoherent, due to 
their fractionalization into gapped orbital excitations, gapless spin and gapless charge modes.
The resulting physics is a simplified 
version of a two-leg ladder in which the 
interactions which usually drive the system away from the exactly solvable point are absent~\cite{BalentsFisher1996,LinBalentsFisher1998}.
 
\begin{table}
\begin{tabular}{|l|l r|}
\hline
Moir\'e length @ $\theta = 0.06^\circ$ & $L_\theta$ & 200 nm\cite{polaritons}\\
\hline
Domain wall width @ $u = 200$ meV & $l$ & 50 nm\cite{polaritons}\\
\hline
Distance to closest gate & $d$ & 25 nm\cite{aharonov}\\
\hline
Minimal distance to NV center & $r$ & 5 nm~\cite{TaylorLukin2008}\\ 
\hline \hline 
Dielectric constant of substrate& $\epsilon$ & 3.2\cite{aharonov}\\
\hline \hline
Velocity of chiral modes & $v$ & 10$^6$ m/s~\cite{mcdonald}\\
\hline
Interaction constant &$g$ & $3 d/(\epsilon l) \times 10^7$ m/s\\
\hline \hline
Displacement field & $u$ & 100 meV\cite{polaritons} \\
\hline
Interlayer hopping (AB regions) & $w$ & 100 meV\cite{BistritzerMacDonald2011}\\
\hline
Dynamical mass gap & $M \sim$ & $w e^{- 2\pi v/g} \sim 20$ meV \\
\hline
Josephson coupling & $\lambda_0$ & 1 meV\\
\hline
Critical temperature @ $\theta = 0.06^\circ$& $T_c$ &0.03 meV\\
\hline
Kondo temperature & $T_K$ & 5 meV~\cite{JiangAndrei2018} \\
\hline
Decay rate of NV spins @ $T = 1$K &$1/T_1$ & $10^{-10}$ meV\\
\hline
\end{tabular}
\caption{Summary of experimental scales in small angle twisted bilayer graphene (see Appendix~\ref{app:EnergyScales} for details).}
\label{tab:Scales}
\end{table}

\section{Experimental probes of fractionalization}
\label{sec:Exp}
\subsection{Tunneling density of states.}  
The most direct
probe of this physics is the local
density of states (LDOS). At temperatures and energies much below the
orbital gap, the LDOS is exponentially suppressed. This is because the leading contribution to the electronic 
Green's function at coinciding space points {requires the creation of
a soliton with energy $E(\theta) = M
\cosh(\theta)$ where $\theta$ is the rapidity.}  The matrix element
for the emission of a soliton with rapidity $\theta$ is fixed by 
Lorentz invariance to be $\exp(S\theta)$ where S is the Lorentz spin,
see Appendix~\ref{app:DOS} for details. In our case $S = 3/16 = 1/2 - 5/16$
(5/16 being the Lorentz spin of the gapless part). The
gapped orbital sector thus contributes a factor $\int d\theta
\exp(2S\theta)\exp[- \vert \tau\vert E(\theta)] = K_{3/8}(M \vert
\tau\vert)$ to the Green's function ($K_n(x)$ denotes the modified Bessel function of the second kind). We note that this result 
is uniquely determined 
by the Lorentz invariance and the scaling dimension of
the operators in the gapless sector. The result for {the total
electronic Green's function is thus} 
\begin{eqnarray}
  G(\tau,x=0) &=& \frac{Z\mbox{sign}(\tau)}{|\tau|^{5/8}}K_{3/8}(M{\vert \tau \vert}) 
 \end{eqnarray}
 where $Z$ is a non universal factor and the term $\sim |\tau|^{-5/8}$ stems from the gapless sector at $K = 1$. After the Fourier transform and analytic continuation we obtain the LDOS:
  \begin{eqnarray}
  \rho(\Omega) \sim \int_0^{\text{arcosh}(\Omega/M)}\frac{\rd\theta \cosh(3\theta/8)}{(\Omega - M\cosh\theta)^{3/8}}. 
  \end{eqnarray}
   The result of integration {and its asymptotic behavior $\rho(\Omega){\sim} (\Omega- M)^{1/8}$} are represented in Fig.~\ref{fig:summary} {\bf b}.

\subsection{Magnetic impurities.} 

One spectacular result,
 unique to the stripes in twisted bilayer graphene
relates to their interaction 
with magnetic impurities which may be generated by vacancies. A vacancy produces broken $\sigma$-orbitals; two of those hybridize  leaving a dangling bond which hosts an unpaired electron.  In flat graphene the corresponding wave function is orthogonal to the $\pi$-orbitals and there is no Kondo screening, but a local curvature lifts this constraint~\cite{NandaSatpathy2012}. The Kondo effect with high Kondo temperature $T_K \sim 68$K was observed in STM experiments conducted on graphene samples deposited on a corrugated substrate~\cite{JiangAndrei2018}. 

In our case the Kondo effect is expected to be quantum critical.  This 
peculiar behavior is protected by the development of an orbital gap. 
Let us consider a domain wall interacting with a single magnetic
moment (Fig.~\ref{fig:summary}{\bf a}). Note that the domain wall width substantially exceeds the atomic scale, see Table~\ref{tab:Scales}, such that impurity spins can be buried inside the domain wall.
The opening of an orbital gap quenches backscattering, so that the
spin density operators $R^+\vec\s L,~L^+\vec\s R$ develop short range
correlations and become irrelevant.  A localized spin then interacts exclusively with spin currents  of the domain wall
\begin{eqnarray} H_{K} = J_K
\sum_a[{J}_R^a(0) + {J}_L^a(0)]\hat{S}^a. \label{eq:Kondo} 
\end{eqnarray} 
Each
spin current belongs to an su$_2$(2) Kac-Moody algebra, an their local
sum belongs to the su$_4$(2) algebra. This is the algebra of a
four-channel Kondo (4CK) model, also 
equivalent to the $M = 3$ topological Kondo
effect~\cite{BeriCooper2012,AltlandEgger2013,altland}, see Appendix~\ref{app:Kondo}. The physics can
be explicitly revealed 
using either the Majorana representation of SU$_2$(2), or 
by treating the anisotropic 4CK problem using Abelian
bosonization at the Toulouse
limit~\cite{FabrizioGogolin1994,Tsvelik1995}.  Thus 
an impurity positioned inside a domain will develop a 4CK effect. 
This physics will develop provided 
the moir\'e length $L_\theta$ exceeds the size of the
Kondo screening cloud $\hbar v/k_B T_K \sim 100$ nm, as realized experimentally
in Ref.~\onlinecite{polaritons}, see Table~\ref{tab:Scales}.

 Amongst the distinct quantum critical properties of the 4CK effect, the residual ground state 
 entropy $S_{\rm imp} = k_{\rm B} \ln(\sqrt{3})$~\cite{Tsvelik1985,AffleckLudwig1991b} signals an emergent 
zero energy ${\mathbb Z_3}$ parafermion mode located at the impurity site. Unlike the charge Kondo effect~\cite{IftikharSimon2018}, where
 transport experiments can probe
 fractionalization~\cite{LandauSela2018}, the situation in graphene 
requires a direct coupling to the spin sector. The setup of TBG
networks, which also contains unscreened impurity spins in the AB/BA
insulating regions necessitates a local probe of the spins 
located on the wires of the network.
 
A promising way to probe the magnetism of spins on a wire network is
to use the energy relaxation rate $1/T_{1}$ of an  NV center spin qubit.
When exposed to a fluctuating magnetic field, this relaxation rate 
is given by~\cite{RodriguezNievaDemler2018}
\begin{eqnarray}
\frac{1}{T_1} = \frac{(g \mu_B)^2}{2 \hbar^2} \coth\left (\frac{\hbar
\Omega}{2 k_B T}\right) \text{ Im}[C_B^R(\Omega)], 
\end{eqnarray}
where $\Omega \ll k_B T/\hbar$ is the frequency of the qubit
(typically GHz) and $C_B^R(\Omega)$ is the retarded Green's function
of local magnetic field $\v B(\v r, t)$ at the qubit position. This
field is a combination of the dipole field of the localized moment
and the fields resulting from currents in the domain wall. 
The
orientation of the intrinsic polarization axis of the qubit permits a
separate measurement of relaxation due to both charge and the spin currents in
the wire. Spin currents give rise to a relaxation
~\cite{RodriguezNievaDemler2018}
\begin{equation}
\left . \frac{1}{T_1} \right \vert_{\rm wire} \sim {\frac{\mu_0^2 g^2 \mu_B^4}{\hbar r^4} \frac{k_B T}{\hbar^2 v^2}} \sim g^2 \times 0.2 \text{mHz}
\end{equation}
at $T = 1$K and $r = 5$nm. 
Near a k-channel Kondo impurity, the critical spin-fluctuations are the dominant channel of relaxation (see~Appendix~\ref{app:Kondo} for details), as their contribution diverges with a characteristic power, see Fig.~\ref{fig:summary} {\bf d}
\begin{equation}
\left .\frac{1}{T_1} \right \vert_{\rm spin} = \mathcal C_\Delta {\frac{\mu_0^2 g^2 \mu_B^4}{\hbar r^6} \frac{1}{k_B T_K}} \left (\frac{T}{T_K} \right)^{2\Delta - 1}.
\end{equation}
Here, $\Delta = 2/(2+k)$ and $\mathcal C_\Delta \sim 1$ is weakly $\Delta$ dependent. For TBG with $T_K = 68$K and $k =4$, the prefactor is $g^2 \times 8$ Hz at $r = 5$nm. 
Importantly, $1/T_1$ diverges with the same power as
the low-T spin susceptibility \cite{Tsvelik84, AffleckLudwig1991} but is a local observable
(at cryogenic temperatures, we expect decay rates of several $100$ Hz which is observable~\cite{BarGill2013} particularly in isotopically pure diamond~\cite{Wrachtrup2009}). While we have discussed the response of a single impurity, it is experimentally favorable to study a collection of Kondo impurities to increase the signal. We highlight that the characteristic power law $T^{-1/3}$ persists when we add up spin-relaxation times $1/T_1$ of a collection of impurity spins with arbitrary distribution of Kondo temperatures.

 As a second experimental probe, we consider two-dimensional
non-linear spectroscopy of the Kondo impurity. The quantum critical
nature of multichannel Kondo effect is reflected in the nontrivial
behavior of the multi-time correlation functions and may be
accessible THz
technology spectroscopy ~\cite{KuehnHey2011,WernerElsaesser2013,WanArmitage2019,ChoiKim2020}, by measuring the
%
%
%
response to pulses of magnetic field applied at particular times
$t_1,...t_{N_1}$. For the multichannel Kondo effect the three-point
function of impurity spins contains signatures of the non-Abelian nature of the QCP~\cite{AffleckLudwig1991,altland}
\begin{equation}
\la\la S^a(t_1)S^b(t_2)S^c(t_3)\ra\ra \sim \epsilon^{abc}W(t_{12})W(t_{13})W(t_{23}),  
\end{equation}
where $W(t) = [{\pi T}/{|\sinh(\pi T t)|}]^{1/3}.$

\section{Coupling between domain walls.} 
\label{sec:2DSC}
We now
re-instate the coupling between the domain walls ("wires").  
In view of the single particle gap $M$, the only operators which can couple are superconducting order parameters~\cite{xu} (see Appendix~\ref{app:OrderParameter} for details):
  \begin{equation}
\hat \Delta_{\substack{p, p'\\\sigma, \sigma'}} = R_{p \sigma} L_{p' \sigma'} \sim  e^{i \sqrt{\pi} \Theta_c}(\tau_y)_{pp'}[\sigma_y(\Delta_s + i\vec \sigma\cdot\vec\Delta_t)]_{\sigma, \sigma'}.\label{delta}
  \end{equation}
where 
$\Theta_{c}$ is
 the superconducting phase, a U(1) Gaussian field describing the charge sector, and $\Delta_s + i \vec \sigma \cdot \vec \Delta_t,\;
 (\Delta_s^2 + \vec \Delta_t^2 = 1)$ is the SU$_2$(2) Wess-Zumino matrix field describing the spin sector. 
 The  pairing operators have a scaling dimension $d = 3/8 + 1/4K$
 ($K<1$ corresponds to repulsion).  To estimate the transition
 temperature 
we use a random phase approximation. This essentially describes the linearized mean field theory described
by the Gaussian model:
\begin{eqnarray}
S &=&\sum_j\int \rd\tau\rd\tau' \; \hat\Delta^*_{j,\mu}({\bf r},\tau)G_{0,j}^{-1}({\bf r}-{\bf r}',\tau-\tau')\hat\Delta_{j,\mu}({\bf r}',\tau')\nonumber\\
 &+&\sum_{j \neq k}\int \rd\tau  \;\lambda_\mu \hat\Delta_{j,\mu}^*({\bf r},\tau) \hat\Delta_{k,\mu}({\bf r},\tau). \label{eq:NetworkAction}
 \end{eqnarray}
Here, the propagator is $G_{0,j}({\bf r},\tau) = {(\pi T/M)^{2d}}/{[\sin^2(\pi T\tau) + \sinh^2(\pi TL_{\theta}{\bf r\cdot e}_j/v)]}^d$, where
 ${\bf e}_j$ are unit vectors of the triangular lattice and $L_\theta$ is
 the size of its unit cell. The sum over sites of the triangular lattice, spin projections $\mu = 0,1,2,3$ of the order parameter field $\hat\Delta_{j,\mu} = \text{tr}^\sigma[\hat \Delta_j \sigma_\mu (i \sigma_y)]$ are implied. 
For $d <1$ ($K > 2/5$) the coupling is relevant and will lead to a transition, at least on the mean field level. 
However, in a 2D the system only an Abelian symmetry can be spontaneously broken at finite T. This means that if the SU(2) symmetry is not broken by anisotropy {of Josephson energies $\lambda_0 \neq \lambda_{1,2,3}$}, the phase transition can occur only at $T=0$. On the other hand in the presence of anisotropy the Ising variables {$\Delta_s$} can order.

Variation of the action with respect to $\Delta^*({\bf r},\tau)$ yields linear mean field equations (see Appendix~\ref{app:Tc}) which are solved by the condition $\lambda_0 \tilde G(\omega = 0, q = 0) = -1/2$. Here $\tilde G(\omega = 0,q = 0) = (\pi T/M)^{2d-2} v/(L_\theta M^2)$ is the integral of the propagator over space-time. (A constant of order unity has been dropped.) Thus,
for $\lambda_0 <0$ we find the mean-field transition temperature for s-wave singlet pairing to be
\begin{equation} \label{eq:Tc}
T_c = M \left (\frac{v \vert\lambda_0 \vert}{M^2L_\theta}\right)^{\frac{1}{2-2d}} = M \left (\frac{v \vert \lambda_0 \vert}{M^2L_\theta}\right)^{\frac{1}{5/4 - 1/2K}}.
\end{equation}
This solution displays equal $\vert \Delta \vert$ on all wires. Using realistic experimental scales, Tab~\ref{tab:Scales}, we obtain $T_c \sim 0.3$ K at $\theta = 0.06^\circ$. 

The focus of our work is to study the network of charge rivers in small angle TBG, and at present it is not clear whether this state is adiabatically connected to the physics of magic-angle twisted bilayer graphene (near $\theta \approx 1^\circ$). Still, it is interesting to extrapolate our result to the magic angle by means of the power law $T_c \propto \theta^{4/3}$. We obtain $T_c \vert _{\theta = 1^\circ} \sim 12$ K, which is about 4 times larger than observed in present day experiments~\cite{LuEfetov2019} (on samples which display a disordered network of normal and superconducting puddles).

The treatment of smooth spatial fluctuations (i.e. the finite momentum term in $\tilde G(0,q)$) on top of the mean field solution allows to incorporate a weak magnetic field and to estimate the upper critical magnetic field near $T_c$, which is  
$eB_{c2} = (2-2d) (1-T/T_c) v^2/T_c^2$. We mention in passing, that for repulsive $\lambda_0 >0$, we also find a (inhomogeneous) mean field solution, but at a temperature which is $2^{-1/(2-2d)}$ smaller. 

\section{Summary and conclusion.} {In summary, we have presented an extensive study of stripe superconductivity in small angle twisted bilayer graphene. We presented a microscopic derivation of coupling constants, which are easy plane anisotropic and drive each wire into a fractionalized, critical $U_4(1) \times SU_2(2)$ state (marked $c = 5/2$ in Fig.~\ref{fig:summary} {\bf b}). We outlined several experimental probes to experimentally verify this theory. In addition to tunneling density of states,~Fig.~\ref{fig:summary} {\bf c}, we studied the interplay of localized spin impurities with the fractionalized quasiparticles and demonstrated the stable emergence of a 4-channel Kondo effect. We derived the response of single-spin quantum sensors to multichannel Kondo impurities, specifically in noise-magnetometry, Fig.~\ref{fig:summary} {\bf d}, and two-dimensional non-linear spectroscopy. Finally, we considered the effect of coupling the wires and derived the mean field transition temperature and critical field toward a 2D singlet s-wave superconductor, plotted in Fig.~\ref{fig:summary} {\bf b}. }

We conclude with an outlook on the parallel to the $x = 1/8$ anomaly of lanthanum cuprates. First, an experimental confirmation of superconductivity in TBG networks is necessary. While superconductivity was observed at the magic angle $\theta = 1.1^\circ$ even for displacement fields $u \sim 100$ meV~\cite{Yankowitz2019} -- which is a significant energy as compared to the band width and might induce networks -- experimental evidence for networks only exists for resistive samples at smaller angles where lattice relaxation is more efficient. Once this intermediate goal is achieved, more theoretical and experimental work is needed to probe both superconducting and parent states. A fascinating question for such studies regards the Hall response, which was shown to be vanishing in normal state $x = 1/8$ LBCO~\cite{tranquada} and places stripe materials in the broader context of anomalous metals (e.g. disordered 2D superconductors~\cite{KapitulnikSpivak2019}). Leaving details to the future, we mention that vanishing Hall response in resistive network TBG may arise in the regime when the coupling $\lambda_\mu$ is irrelevant or due to the emergent particle-hole symmetry of the quasi-one dimensional model.

\section{Acknowledgments}

It is a pleasure to acknowledge useful discussions with Eva Andrei, Yang-Zhi Chou, Dmitry Efimkin, Yuhang Jiang, Jan Jeske and Jinhai Mao. Funding: PC and EJK were supported by the U.S. Department of Energy, Office of Basic Energy Sciences, under contract No. {DE-FG02-99ER45790}, AMT was supported by the U.S. Department of Energy, Office of Basic Energy Sciences, under contract No. {DE-SC0012704}.
  
\appendix

\section{Microscopic derivation of interaction matrix elements.}

\label{app:MicroDerivation}

In this appendix we derive Eqs.~\eqref{H0},\eqref{currents} microscopically. Contrary to the main text, we use the notation $q = \pm 1$ for the orbital quantum number (in order to avoid confusing the orbital quantum number with the momentum $p$).

\subsection{Wave function at the domain wall}
\label{app:wavefunctions}

We briefly summarize the main results of Efimkin and MacDonald~\cite{mcdonald} to introduce the notation for the next section.
Following the procedure outlined there, we obtain the effective Hamiltonian for the $K$ valley (to leading order in small twist angles $\theta \ll 1$)
\begin{widetext}
\begin{eqnarray}
H_D &=& \left (\begin{array}{cc}
v_0 p - u & -\delta_-(\v r) \cos(\phi_{\v p} - \Theta) - i \delta_+(\v r) \sin(\phi_{\v p} - \Theta) \\ 
-\delta_-(\v r) \cos(\phi_{\v p} - \Theta) + i \delta_+(\v r) \sin(\phi_{\v p} - \Theta) & -(v_0 p - u) \end{array} \right ) \notag \\
&\simeq & \pm \left (\begin{array}{cc}
v_0 \hat p_\perp &  (-\delta_-(\v r) \mp i v_{\Vert} \hat p_{\Vert}) \\ 
(-\delta_-(\v r) \pm i v_{\Vert} \hat p_{\Vert}) & -v_0 \hat p_\perp
\end{array} \right ).  \label{eq:LowEnergy}
\end{eqnarray} 
\end{widetext}
Here, the two-by-two structure corresponds to the two graphene monolayers, while the sublattice degree of freedom has been projected out (states at energy $2 u$ were neglected). For simplicity, we follow the notation of Ref.~\onlinecite{mcdonald}, according to which $\delta_\pm(\v r) = [\vert T_{AB}(\v r)\vert \pm \vert T_{BA}(\v r)\vert]/2$ (not to be confused with the delta function) is defined by means of the interlayer hopping elements of the Bistritzer-MacDonald model of twisted bilayer graphene~\cite{BistritzerMacDonald2011}. Moreover, $p_x + i p_y = p e^{i \phi_{\v p}}$ and $\Theta = \Theta(\v r) = \text{arg}[\sqrt{T_{AB}^*(\v r)T_{BA}(\v r)}]$.
In the second line, we have expanded near the two minima of the dispersion $\phi_{\v p} = \Theta$ and $\phi_{\v p} = \Theta + \pi$ (see Fig.~\ref{fig:orbital} {\bf a} of the main text), leading to two $2\times 2$ Hamiltonians, with opposite prefactor and $v_\Vert = \delta_+/p_u$ (here $\delta_+(\v r)$ is evaluated on a domain wall). Note that $\Theta = 0$ near the domain wall $x =0$, see Fig.~\ref{fig:orbital} {\bf a} of the main text and Fig.~2 in Ref.~\cite{mcdonald} and that near the two relevant momenta the prefactor of $\delta_{-}$ is opposite. Moreover, the positions of the gap closings are opposite to each other on the circle $p = p_u$ and thus the perpendicular direction is also opposite, which allows to extract the overall $\pm$ sign in the second line. In this line we also reinstated the operator nature of momenta (denoted by a hat). In particular, $\delta_+ \phi_{\v p} =\pm v_\Vert \hat p_\Vert$ (the parallel momentum is related to the phase with opposite sign on opposite sides of the Fermi circle).
In Fig.~\ref{fig:orbital} of the main text, $\hat p_\perp = \hat p_x$ and $\hat p_\Vert = \hat p_y$.

Using a standard Jackiw-Rebbi calculation there are two chiral dispersing states (with index $q = \pm 1$) per spin moving upstream at the $K$ and downstream at the $K'$ valleys
\begin{subequations}
\begin{align}
\psi_{K, q}^{(p_\Vert)}(\v r) &= e^{i p_\Vert \hat e_\Vert \cdot \v r + i q p_u \hat e_\perp \cdot \v r + i \v K \cdot \v r} e^{-\int^{r_{\perp}} \delta_-(r') dr'/v_0} \left( \begin{array}{c}1 \\i \end{array}\right) \label{eq:EfimkinMacDonaldWFK}\\
\psi_{K', q}^{(p_\Vert)}(\v r) &= e^{-i p_\Vert \hat e_\Vert \cdot \v r - i q p_u \hat e_\perp \cdot \v r - i \v K \cdot \v r} e^{-\int^{r_{\perp}} \delta_-(r') dr'/v_0} \left( \begin{array}{c}1 \\-i \end{array}\right) \label{eq:EfimkinMacDonaldWFKp}
\end{align}
\label{eq:EfimkinMacDonaldWF}
\end{subequations}
Here, the unit vectors $\hat e_\Vert (\hat e_{\perp})$ point along (perpendicularly) to the domain wall, respectively. States at the $K'$ valley are obtained by employing time reversion (i.e. complex conjugation).

This concludes the derivation of Eq.~\eqref{H0} of the main text, where we use field operators $R_{q,\sigma},R^+_{q,\sigma}$  [$L_{q,\sigma}, L^+_{q,\sigma}$] to annihilate/create the modes presented in Eq.~\eqref{eq:EfimkinMacDonaldWFK} [\eqref{eq:EfimkinMacDonaldWFKp}].

\subsection{Bare interaction matrix elements}
Various integrals determine the effective interactions, and we follow the notation of Ref.~\onlinecite{xu} (moreover, we use the notation $\psi_{P, q}(\v r) = \psi_{p_\Vert = 0, P = K/K', q = \pm 1}(\v r)$). Note however, that the effective dispersion, Fig.~\ref{fig:orbital} of the main text differs from Ref.~\onlinecite{xu}, who moreover did not study the microscopic details of the spinor. 
The spinor structure of the Efimkin-MacDonald wave functions Eq.~\eqref{eq:EfimkinMacDonaldWF} implies
\begin{equation}
\psi_{P, q}^\dagger(\v r) \psi_{P', q'} (\v r')  \propto \delta_{PP'}. \label{eq:ExactOrthogonality}
\end{equation}
The direct interaction is 
\begin{eqnarray}
E_0 &=& \int d^2x_1 d^2x_2 \vert \psi_{P, q} (\v x_1) \vert^2 V_{\v x_1, \v x_2}\vert \psi_{P', q'} (\v x_2) \vert^2 \notag \\
&=& \int d^2x_1 d^2x_2 F(x_{1, \perp}) F(x_{2, \perp}) V_{\v x_1, \v x_2},
\end{eqnarray}
here, $F(r_{\perp}) = 2e^{- 2 \int^{r_\perp} \delta_-(r') dr'/v_0}$.

Next, we consider exchange integrals. Following Eq.~\eqref{eq:ExactOrthogonality}, it follows that the  exchange integrals denoted $I_{ex,2}, \dots, I_{ex,5}$ in Ref.~\onlinecite{xu} are exactly zero (which is a stronger statement than reported in the literature~\cite{xu} where it was argued that these integrals should be small due to the large transferred momentum).
The only exchange interaction integrals which do contribute are of the form 
\begin{widetext}
\begin{eqnarray}
E_{q_1,q_1', q_2, q_2'} &=& \int d^2x_1 d^2x_2 \psi_{P, q_1}^\dagger (\v x_1) \psi_{P, q_1'}(\v x_1)  V_{\v x_1, \v x_2} \psi^\dagger_{P', q_2} (\v x_2)\psi_{P',-q_2'} (\v x_2)\notag \\
&=& \int d^2x_1 d^2x_2 F(x_{1, \perp})F(x_{2, \perp}) V_{\v x_1, \v x_2} e^{-i p_u [(q_1 - q_1') x_{1, \perp} + (q_2 - q_2') x_{2, \perp}]} \label{eq:InteractionMatrix}
\end{eqnarray}
\end{widetext}
For diagrammatic illustration of this equation, see Fig.~\ref{fig:orbital} {\bf c}. It also illustrates that the spin index is conserved at the vertices.

\subsubsection{Evaluation for contact interaction}
We evaluate the interaction integrals for contact interaction (valid when the gates are so close that the long-range Coulomb interaction is screened to a length scale shorter than the moir\'e lattice constant). We further assume a simple Gaussian envelope of the bound states $F(x) = e^{-x^2/l^2}$ where $l/\sqrt{2}$ is the width of the domain wall. Then the dimensionless integrals over transverse coordinates are 
\begin{eqnarray}
I_{q_1,q_1', q_2, q_2'}&=& \sqrt{2/\pi}\int dx e^{- 2x^2/l^2}e^{-i p_u(q_1 - q_1' + q_2 - q_2') x} \notag \\
 &=& e^{- (q_1 - q_1' + q_2 - q_2')^2 p_u^2l^2/8}.
\end{eqnarray}
We switch back to the second quantized representation introduced at the end of Appendix~\ref{app:wavefunctions} in the presentation of the following effective Hamiltonian:
\begin{eqnarray}
\mathcal H_{\rm int} &=& g_0 [\sum_{q = \pm 1} (R^+_q R_q + L^+_q L_q) ]^2 \notag \\
&+& g_1 \sum_{q = \pm 1} [(R^+_q R_{-q} + L^+_q L_{-q} )(R^+_{-q} R_{q} + L^+_{-q} L_{q}) ]\notag\\
&+& 2g_2 \sum_{q, q' = \pm 1} [(R^+_q R_{-q} + L^+_q L_{-q} )(R^+_{q'} R_{q'} + L^+_{q'} L_{q'}) ] ]\notag\\
&+& g_3 \sum_{q = \pm 1} [(R^+_q R_{-q} + L^+_q L_{-q} )(R^+_q R_{-q} + L^+_q L_{-q} ) ],
\end{eqnarray}
where $g_1/g_0 = 1$, $g_2/g_0 = e^{- (p_u l)^2/2}$ and $g_3/g_0 = e^{- 2 (p_u l)^2}$. Here, a spinor notation and implicit contraction of spin indices, e.g. $R^+_qR_q \equiv \sum_\sigma R^+_{q,\sigma}R_{q,\sigma}$, are employed.

For the emergence of a dynamical gap, only non-chiral terms containing both right and left currents (i.e. $J_R J_L$ terms) are important and kept.
Furthermore we can use $R^+_q R_{- q} = R^+ \frac{\tau_1 + i q \tau_2}{2} R$, where $R$ is now a 4-spinor in spin and orbital space. We obtain
\begin{eqnarray}
\mathcal H_{\rm int} &=& 2g_0 (R^+ R) (L^+ L) \notag\\
&+& g_1 \sum_{A = x,y} (R^+ \tau_A R)  (L^+ \tau_A L) \notag\\
&+&2 g_2 [(R^+ \tau_1 R) (L^+ L) + (L^+ \tau_1 L) (R^+ R)] \notag\\
&-& g_3  (R^+ \tau_x R) (L^+ \tau_x L)\notag \\
&+& g_3  (R^+ \tau_y R) (L^+ \tau_y L).
\end{eqnarray}

When $p_u l \rightarrow \infty$ (this is the experimentally relevant case, see Table~\ref{tab:Scales}), $g_{2,3} = 0$ and a U(1) symmetry emerges, which is the origin of Eq.~\eqref{currents} of the main text. On the other hand, when $p_u l \rightarrow 0$ we obtain $g_0 = g_1 = g_2 = g_3$, i.e. 
\begin{eqnarray}
\mathcal H_{\rm int} &=& 2g_0 (R^+ [1 + \tau_1] R)(L^+ [1 + \tau_1] L).
\end{eqnarray} 

This derivation of interaction matrix elements is based on the wave functions of Ref.~\onlinecite{mcdonald} and is formally valid for $u \gg w \sim \text{max}(\vert T_{\rm AB} \vert,\vert T_{\rm BA} \vert)$. Crucially, it predicts degenerate single particle states $q = \pm 1$, see Fig.~\ref{fig:orbital}. The degeneracy is lifted as $u$ decreases (see e.g. Ref.~\onlinecite{polaritons}).\cite{Efimkinprivate} For the structure of interaction matrix elements $g_{0,1}$, the important ingredient is only the spinor structure in Eq.~\eqref{eq:EfimkinMacDonaldWF}, which persists to smaller $u$. Therefore, we expect the nature of the interactions, Eq.~\eqref{currents}, to hold at realistic $u \sim w$.

\section{Abelian bosonization and many-body physics}
\label{app:Bosonization}

This appendix contains technical details on the many-body physics of a single domain wall and the coupling of rivers of charge. As in Appendix~\ref{app:MicroDerivation}, we use the notation $q = \pm 1$ for the orbital quantum number (in order to avoid confusing the orbital quantum number with the momentum $p$).

\subsection{Abelian bosonization and refermionization}

\subsubsection{Bosonization rules and Klein factors}

The bosonization rules are ($\sigma = \pm 1, {q} = \pm 1$, {and we use} $\Phi = \varphi +\bar\varphi, ~~ \Theta = \varphi -\bar\varphi$):
\begin{align}
R_{{q},\sigma} &= \frac{\xi_{{q}\sigma}}{\sqrt{2\pi a_0}}\exp[i\sqrt{\pi}(\varphi_c + {q}\varphi_f + \sigma\varphi_s + {q}\sigma\varphi_{sf})], \\
L_{{q},\sigma} &= \frac{ \xi_{{q}\sigma}}{\sqrt{2\pi a_0}}\exp[-i\sqrt{\pi}(\bar\varphi_c + {q}\bar\varphi_f + \sigma\bar\varphi_s + {q}\sigma\bar\varphi_{sf})],
\end{align}
where $\{\xi_a,\xi_b\} = 2\delta_{ab}$ are Klein factors. 

To choose one irreducible representation we have to impose a constraint, for instance  $\xi_{+\uparrow}\xi_{-\downarrow}\xi_{-\uparrow}\xi_{+\downarrow} =-1$. 
Some consequences which follow from this convention are
\begin{align}\label{eq:Kleinidentities}
\xi_{q \sigma} \xi_{\bar q \sigma} &= q \xi_{+ \uparrow} \xi_{- \uparrow} &
\xi_{q \sigma} \xi_{\bar q \bar \sigma} &= \sigma \xi_{+ \uparrow} \xi_{- \downarrow}, &
\xi_{q \sigma} \xi_{q \bar \sigma} &= q\sigma \xi_{+ \uparrow} \xi_{+ \downarrow}.
\end{align}
We can choose, for example $\xi_{+\uparrow} = \alpha^x\otimes\beta^z, ~\xi_{+,\downarrow} = \mathbf 1\otimes\beta^y, ~ \xi_{-\downarrow} = \mathbf 1 \otimes\beta^x, ~\xi_{-,\uparrow} = \alpha^y \otimes\beta^z$ and project everything using the constraint $\alpha^z\beta^z|\text{Phys}\ra = -|\text{Phys}\ra$ ($\alpha^\mu$, $\beta^\mu$ are Pauli matrices).

\subsubsection{Currents}

We now bosonize and subsequently refermionize spin and orbital currents. We consider right moving currents (analogous equations hold for $R \rightarrow L$)
\begin{subequations}
\begin{eqnarray}
R^+ \sigma_+ R &=& \frac{1}{2\pi a_0}\sum_q \xi_{q \uparrow} \xi_{q \downarrow} e^{-i \sqrt{4\pi} (\phi_s + q \phi_{\rm sf})} \notag \\&=& \frac{\xi_{+ \uparrow} \xi_{+ \downarrow} }{2\pi a_0}\sum_q q e^{-i \sqrt{4\pi} (\phi_s + q \phi_{\rm sf})} \notag \\&=& (R_{sf}^\dagger-R_{sf})R_s^\dagger, \\
R^+ \sigma_- R &=& \frac{1}{2\pi a_0}\sum_q \xi_{q \downarrow} \xi_{q \uparrow} e^{i \sqrt{4\pi} (\phi_s + q \phi_{\rm sf})} \notag \\&=&-   \frac{\xi_{+ \uparrow}\xi_{+ \downarrow}}{2\pi a_0}\sum_q q  e^{i \sqrt{4\pi} (\phi_s + q \phi_{\rm sf})} \notag \\
&=& (R_{sf}^\dagger-R_{sf})R_s, \\
R^+ \sigma_3 R &=& 2R^+_s R_s\\
R^+ \tau_+ R &=& \frac{1}{2\pi a_0}\sum_\sigma \xi_{+ \sigma} \xi_{- \sigma} e^{-i \sqrt{4\pi} (\phi_f + \sigma \phi_{\rm sf})} \notag \\&=& \frac{ \xi_{+ \uparrow} \xi_{- \uparrow}}{2\pi a_0}\sum_\sigma e^{-i \sqrt{4\pi} (\phi_f + \sigma \phi_{\rm sf})} \notag \\
&=&  (R_{sf}+R^+_{sf})R_f^\dagger\\
R^+ \tau_- R &=& \frac{1}{2\pi a_0}\sum_\sigma \xi_{- \sigma} \xi_{+ \sigma} e^{i \sqrt{4\pi} (\phi_f + \sigma \phi_{\rm sf})} \notag \\
&=& -\frac{ \xi_{+ \uparrow} \xi_{- \uparrow}}{2\pi a_0}\sum_\sigma  e^{i \sqrt{4\pi} (\phi_f + \sigma \phi_{\rm sf})} \notag \\
&=& -(R^+_{sf} + R_{sf})R_f \\
R^+ \tau_z R &=& 2R^+_f R_f.
\end{eqnarray}
\end{subequations}
Here, we used the identities~\eqref{eq:Kleinidentities} and absorbed $\xi_{+\uparrow}$ into $R_{sf}$, $\xi_{+\downarrow}$ into $R_s$, $\xi_{- \uparrow}$ into $R_f$.
We express the Dirac fermions three $R_f, R_s, R_{sf}$ by two triplets of Majorana fermions
\begin{align} \label{eq:Majoranas}
R_s &= \frac{-\chi^R_2 -i \chi_1^R}{2}, &R_f &= \frac{\zeta_x^R -i \zeta_y^R}{2}, &R_{sf} &= \frac{\zeta^R_z +i \chi_3^R}{2},
\end{align}
such that 
\begin{subequations}
\begin{align}
R^+ \sigma_1 R &= (R^+_{sf} - R_{sf})(R^+_s + R_s)=  -i \chi_2^R \chi_3^R, \notag \\
R^+ \sigma_2 R &=(R^+_{sf} - R_{sf})(-iR^+_s + i R_s) = -i \chi_3^R \chi_1^R, \notag \\
R^+ \sigma_3 R &= -i \chi_1^R \chi_2^R,\\
R^+ \tau_x R &=(R_{sf}^\dagger + R_{sf})(R_f^\dagger - R_f)= - i \zeta_y^R \zeta_z^R, \notag \\
R^+ \tau_y R &= -i(R_{sf}^\dagger + R_{sf})(R_f^\dagger + R_f)= - i \zeta_z^R\zeta_x^R,\notag \\
R^+ \tau_z R &=- i \zeta_x^R \zeta_y^R.
\end{align}
\end{subequations}
Using this representation, it is evident that the interaction, Eq.~\eqref{currents} of the main text, only affects $\zeta$ fields and the currents entering the interaction can be expressed as $J^A =- i \epsilon^{ABC} \zeta_B \zeta_C/2$. We can explicitly rewrite Eq.~\eqref{currents} as

\begin{eqnarray}
{H_{\rm int}}& =& g(\zeta^z_R\zeta^z_L)[(\zeta^x_R\zeta^x_L) + (\zeta^y_R\zeta^y_L)] +g'(\zeta^x_R\zeta^x_L)(\zeta^y_R\zeta^y_L)  \nonumber\\
&=& \frac{g'}{2\pi}(\p_{\mu}\Phi_f)^2 - \ri g \cos(\sqrt{4\pi}\Phi_f)\zeta^z_R\zeta^z_L, 
\end{eqnarray}
where $g'(0) << g(0)$. All couplings are positive and scale to strong coupling. The vacuum corresponds to $\Phi_f =0, \la\mu^z\ra \neq 0$ or $\Phi_f = \sqrt\pi/2, \la\sigma^z\ra \neq 0$, where $\sigma, \mu$ correspond to order and disorder operators of the Ising theory. In the Ising model basis it corresponds to either all $\mu$'s or all $\sigma$'s having nonzero vacuum averages. 

\subsection{Expression in terms of Ising fields and superconducting order parameter}

We use the following dictionary to map~\cite{TsvelikQFT} operators to primary fields of six separate transverse field Ising models denoted by $x,y,z$ and $1,2,3$ in correspondence to the Majorana modes introduced in Eq.~\eqref{eq:Majoranas}
\begin{subequations}
\begin{align}
\cos(\sqrt{\pi}\Phi_f) &\sim \mu_x \mu_y, & \sin(\sqrt{\pi}\Phi_f) &\sim \sigma_x \sigma_y, \\
\cos(\sqrt{\pi}\Theta_f) &\sim \mu_x \sigma_y, & \sin(\sqrt{\pi}\Theta_f) &\sim \sigma_x \mu_y, \\
\cos(\sqrt{\pi}\Phi_s) &\sim \mu_1 \mu_2, & \sin(\sqrt{\pi}\Phi_s) &\sim \sigma_1 \sigma_2, \\
\cos(\sqrt{\pi}\Theta_s) &\sim \mu_1 \sigma_2, & \sin(\sqrt{\pi}\Theta_s) &\sim \sigma_1 \mu_2 ,\\
\cos(\sqrt{\pi}\Phi_{sf}) &\sim \mu_z \mu_3, & \sin(\sqrt{\pi}\Phi_{sf}) &\sim \sigma_z \sigma_3, \\
\cos(\sqrt{\pi}\Theta_{sf}) &\sim {\sigma_z \mu_3}, & \sin(\sqrt{\pi}\Theta_{sf}) &\sim {\mu_z \sigma_3}.
\end{align}
\label{eq:IsingFields}
\end{subequations}

\subsection{Superconducting order parameter}
\label{app:OrderParameter}

The superconducting order parameter is
\begin{align}
\Delta_{\substack{q, q'\\\sigma, \sigma'}} &= R_{q \sigma} L_{q' \sigma'} \notag\\
&\sim \xi_{q \sigma} \xi_{q' \sigma'} e^{i \sqrt{\pi}[\phi_c +\sigma \phi_s + q \phi_f + q \sigma \phi_{sf}]}\notag \\
&\times e^{-i \sqrt{\pi}[\bar\phi_c +\sigma'\bar \phi_s + q'\bar \phi_f + q' \sigma'\bar \phi_{sf}]}.
\end{align}
Only $q = - q' \equiv \bar q'$ terms are non-zero once $\Phi_f$ orders. Thus
\begin{widetext}
\begin{align}
\Delta_{\substack{q, q'\\\sigma, \sigma'}} &\sim M^{1/4} e^{i \sqrt{\pi} \Theta_c} \xi_{q \sigma} \xi_{q' \sigma'}\delta_{q \bar q'} \Big \{ \delta_{\sigma \sigma'} \big [ e^{i \sigma \sqrt{\pi} \Theta_s} e^{i \sigma  q \sqrt{\pi} \Phi_{sf}} \big ] + \delta_{\sigma \bar \sigma'} \big [ e^{i \sigma \sqrt{\pi} \Phi_s} e^{i \sigma q \sqrt{\pi} \Theta_{sf}} \big ]\Big \}\notag\\
&= - [\hat \tau_y]_{qq'} M^{3/8} e^{i \sqrt{\pi} \Theta_c}  \left (\begin{array}{cc}
\alpha_z \cos(\sqrt{\pi}\Phi_{\rm sf}) e^{i \sqrt{\pi} \Theta_s} & -\alpha_y \beta_x \sin(\sqrt{\pi}\Theta_{\rm sf}) e^{i \sqrt{\pi} \Phi_s} \\ 
-\alpha_y \beta_x \sin(\sqrt{\pi}\Theta_{\rm sf}) e^{-i \sqrt{\pi} \Phi_s}& \alpha_z \cos(\sqrt{\pi}\Phi_{\rm sf}) e^{-i \sqrt{\pi} \Theta_s}
\end{array}  \right )_{\sigma \sigma'}
\end{align}
\end{widetext}
At the second equality sign, we used that, when $\mu_z$ orders, $\sin(\sqrt{\pi}\Phi_{sf}) = 0,\cos(\sqrt{\pi}\Theta_{sf}) = 0$ and employed the representation of Klein factors introduced in the beginning of Appendix~\ref{app:Bosonization}. 

In summary we thus obtain for the order parameter matrix field
\begin{subequations}
\begin{equation}
\Delta = i \alpha_z \hat \tau_y M^{3/8} e^{i \sqrt{\pi} \Theta_c} \hat g
\end{equation}
where
\begin{eqnarray}
\hat g &=& \alpha_x \beta_x \hat{\mathbf 1}_{ \sigma} \sigma_1 \sigma_2 \sigma_3 + i \hat \sigma_1 \sigma_1 \mu_2 \mu_3 \notag \\
&& - i \hat \sigma_2 \mu_1 \sigma_2 \mu_3 + i \alpha_x \beta_x \hat \sigma_3 \mu_1 \mu_2 \sigma_3.
\end{eqnarray}
\end{subequations}
This identifies the order parameter field with $SU(2)_2$ fields in their representation by order and disorder operators of three Ising theories~\cite{TsvelikQFT} and concludes the derivation of Eq.~\eqref{delta} of the main text. (We there did not dwell on the subtleties of Klein factors.) Note that $\hat \sigma_{1,2,3}$ are spin operators, and $\sigma_{1,2,3}$ the order operators of the three distinct Ising models. (In the main text, we do not explicitly introduce the Ising operators and therefore represent spin operators without a hat.)  

\subsection{Coupled wires and 2D superconductivity}
\label{app:Tc}

\subsubsection{Transition temperature}

This section contains details about the transition temperature of the effective model, Eq.~\eqref{eq:NetworkAction}. 
{Variation of the action with respect to $\Delta^*({\bf r},\tau)$ yields linear mean field equations} {(we only keep the singlet channel)}
 
\begin{equation}
\left (\begin{array}{ccc}
\delta_{x_1,x_1'} & \lambda G^{0}_{x_1 x_2'}   & \lambda G^{0}_{x_1 x_3'} \\ 
\lambda G^{0}_{x_2 x_1'} & \delta_{x_2 x_2'} & \lambda G^{0}_{x_2 x_3'} \\ 
\lambda G^{0}_{x_3 x_1'} & \lambda G^{0}_{x_3 x_2'} & \delta_{x_3 x_3'}
\end{array} \right ) \left (\begin{array}{c}
\Delta_{x_1'} \\ 
\Delta_{x_2'} \\ 
\Delta_{x_3'}
\end{array} \right) = 0.
\end{equation} 
{Here, summation/integration over repeated space time variables $x_i = ({\bf r_i}, \tau)$ is assumed ($x_i$ lives on wires in direction $\hat e_i$). Note that all mean field equations were multiplied by $G_{xx'}$ from the left. We seek a homogeneous, static solution, leading to the mean field condition}
\begin{equation}
1 - 3 (\lambda \tilde G(0,0))^2 + 2(\lambda \tilde G(0,0))^3 =0 \Leftrightarrow \lambda \tilde G(0,0) \in \left \{ -\frac{1}{2},1 \right \}.
\end{equation}
{Here, we introduced the zero frequency limit of the Fourier transform}
\begin{align}
 \tilde  G(0,q)& = \frac{1}{\pi T}\left  (\frac{\pi T}{M} \right)^{2d}\frac{v}{\pi L_\theta T} {\cal F}_d(v/\pi L_\theta T,q); \nonumber\\
 {\cal F}_d(x,q) &= x^{-1}\sum_{n=1}^{\infty}\frac{1}{\pi}\int_0^{\pi} \frac{\rd\tau e^{i q L_\theta  n}}{[\sin^2\tau + \sinh^2(n/x)]^d} \notag\\
 & \stackrel{x\gg 1}{\simeq}  \int_{1/x}^\infty  dy \int_0^\pi \frac{d\tau}{\pi} \frac{e^{i (q l_T) y}}{[\sin^2\tau + \sinh^2(y)]^d}. \label{eq:Fxq}
 \end{align}
Here, $l_T = v/(\pi T)$ is the thermal length. {The integral diverges at the lower bound (UV) if $d >1$ i.e. $K <2/5$ - in the opposite case (corresponding to relevant Josephson coupling) $\mathcal F_d(x,q)$ approaches a constant at $x\rightarrow \infty$. {Then, $\tilde  G(0,q) \sim \left (\pi T/M \right)^{2d - 2} v/M^2L_\theta  (1 + \mathcal C (q l_T)^2)$, as quoted in the main text.}

Using these results for $\lambda <0$ we find the mean-field transition temperature}
\begin{equation}
T_c = M \left (\frac{v \vert\lambda \vert}{M^2L_\theta }\right)^{\frac{1}{2-2d}} = M \left (\frac{v \vert \lambda \vert}{M^2L_\theta }\right)^{\frac{1}{5/4 - 1/2K}}.
\end{equation}
This solution displays finite superconductivity on all wires, which are all in phase. Incidentally, at $\lambda >0$ (repulsion), we also find a solution, but at a temperature which is $2^{1/(2d-2)}$ smaller. However, this solution of the mean field Eq. (14) only has two wires with non-zero superconducting order parameters which are in antiphase. 

Finally, we briefly check the consistency of our approximations, i.e. whether $T_c L_\theta /v  \ll 1$ is satisfied (justifying the expansion in $x \gg 1$ in Eq.~\eqref{eq:Fxq}). We find
\begin{equation}
\frac{T_c L_\theta }{v} = \left [ \frac{\vert \lambda \vert}{M} \left (\frac{ML_\theta }{v} \right)^{1-2d} \right ]^{\frac{1}{2-2d}}  \ll 1 \Leftrightarrow \frac{\vert \lambda \vert}{M} \ll \Big (\underbrace{\frac{v}{ML_\theta }}_{\ll 1} \Big)^{1-2d}.
\end{equation} 
For the most relevant regime $1/2 < d < 1$, the exponent on the right hand side is negative, such that the assumption is met so long as $M$ is the largest scale in the problem. Using the experimental parameters of Table~\ref{tab:Scales}, $\hbar v/L_\theta \sim 3$ meV is indeed substantially smaller than $M \sim 20$ meV and $\vert \lambda\vert/M \sim 1/2$.

\subsubsection{Critical magnetic field}

We now estimate the upper critical field near the mean field $T_c$. To this end, we consider the long-wavelength limit of Eq.~\eqref{eq:NetworkAction}, i.e. we Fourier transform keeping only the zeroth frequency and up to second order in momentum and rescale all fields by $\sqrt{\tilde G(0,0)}$. Thereby we obtain the renormalized free energy
\begin{widetext}
\begin{equation}
F = \int d^2x \vec \Delta^\dagger(\vec x) \left (\begin{array}{ccc}
1 -  l_T^2(\hat e_1 \cdot \vec \nabla)^2 & \tilde \lambda & \tilde \lambda \\ 
\tilde \lambda & 1 -  l_T^2(\hat e_2 \cdot \vec \nabla)^2 & \tilde \lambda \\ 
\tilde \lambda & \tilde \lambda & 1 -  l_T^2(\hat e_3 \cdot \vec \nabla)^2
\end{array}  \right) \vec \Delta(\vec x).
\end{equation}
\end{widetext}
Here, $\tilde \lambda = \lambda \tilde G(0,0)$, a constant has been absorbed into $l_T$, and the long derivative is $-i \vec \nabla = - i \vec \partial - \vec A(\vec x)$. Under the assumption that the instability is isotropic in the space of $\Delta_{1,2,3}$ we obtain project this effective Hamiltonian on the $(1,1,1)$ state and obtain
\begin{equation}
F = \int d^2x \Delta_s^* (1 + 2 \tilde \lambda - l_T^2 \vec \nabla^2) \Delta_s,
\end{equation}
which is the standard expression for the linearized Ginzburg-Landau functional, and leads to the standard linear behavior of $H_{c2}$ near $T_c$ obtained by $eB = -(1+2 \tilde \lambda)/l_T^2$.

\section{Tunneling Density of states}
\label{app:DOS}

In this section, we present details on the derivation of the tunneling density of states. The correlation function of the gapped orbital part of the fermionic operators (represented by an $\mathcal O(x,\tau)$) can be expressed in terms of form factors (see e.g.~\cite{TsvelikQFT,EsslerKonik2005} for a pedagogic review)
\begin{align}
&-\langle 0 \vert \mathcal T \mathcal O^\dagger(0,\tau) \mathcal O(0,0) \vert 0 \rangle = \sum_{N} \sum_{\{\theta_i, a_i \}} e^{- E_{\{\theta_i, a_i \}} \tau} \notag\\
&\times \langle 0 \vert \mathcal O^\dagger(0,0) \vert  \{\theta_i, a_i \}_{i = 1}^N \rangle \langle \{\theta_i, a_i \}_{i = 1}^N \vert \mathcal O(0,0) \vert 0 \rangle.
\end{align}
We have used a resolution of the identity and introduced a symbolic notation $ \vert  \{\theta_i, a_i \}_{i = 1}^N \rangle$ for an $N$ particle state with energy $E_{\{\theta_i, a_i \} } = M \sum_j \cosh(\theta_j)$ and characterized by rapidities $\theta_i$ and isotopic index $a_i$ in the gapped integrable orbital sector. For our purposes it is sufficient to consider a single particle excited state containing one soliton. In this simple case the form factor
\begin{equation}
F_a(\mathcal O^\dagger, \theta) = \langle 0 \vert \mathcal O^\dagger(0,0) \vert \theta, a \rangle
\end{equation}
is determined to be $F_a(\mathcal O^\dagger, \theta) \sim e^{S \theta}$ because of Lorentz invariance, i.e. $F_a(\mathcal O^\dagger, \theta) = e^{S \alpha} F_a(\mathcal O^\dagger, \theta + \alpha)$. The Lorentz spin $S$ can be determined by subtracting the gapless contribution ($S = 5/16$ for our model $\mathcal H[U(1)] + \mathcal H[SU_2(2)]$) from the free fermion case ($S = 1/2$) leading to $S = 3/16$. We thus obtain
\begin{eqnarray}
-\langle 0 \vert \mathcal T \mathcal O^\dagger(0,\tau) \mathcal O(0,0) \vert 0 \rangle &\sim & \int_{-\infty}^\infty d \theta e^{\frac{3 \theta}{8} - M \vert \tau \vert \cosh(\theta)}\notag \\ &=& K_{3/8}(M \vert \tau \vert).
\end{eqnarray}
We remind the reader of the asymptotes $K_{3/8}(x) \sim x^{-3/8} (0<x\ll 1)$ and $K_{3/8}(x) \sim e^{-x}/\sqrt{x} (x\gg 1)$. 

With the help of this expression, we perform a Fourier transformation of the Green's function using
\begin{eqnarray}
\int_0^\infty d\tau \frac{e^{\pm i\omega \tau -M \cosh(\theta)\tau}}{\tau^{5/8}} &=& \frac{1}{(M\cosh(\theta)\mp i \omega)^{3/8}} \notag\\
&&\times \lim_{\Lambda \rightarrow \infty} \int_0^{\Lambda e^{i\phi_\pm}} dx \frac{e^{-x}}{x^{5/8}}, \notag
\end{eqnarray}
where $\phi_\pm = \text{arg}(M\cosh(\theta)\mp i \omega)$. Thus 
\begin{equation}
G(\Omega^+) \sim \int d\theta \cosh(3 \theta/8) \sum_\pm \frac{\pm 1}{(M\cosh(\theta)\mp \Omega^+)^{3/8}}.
\end{equation}
The imaginary part of the tunneling density of states and its asymptotic behavior are
\begin{eqnarray}
 \rho(\Omega) &\sim& \int_0^{\text{arcosh}(\Omega/M)}\frac{\rd\theta \cosh(3\theta/8)}{(\Omega - M\cosh\theta)^{3/8}} \notag \\
 &\simeq& \left . \int_0^{\sqrt{2 x}}\frac{d \theta}{(x^2 - \theta^2)^{3/8}} \right \vert_{x = \sqrt{2 (\Omega/M-1)}}\notag\\
 &\sim& x^{1/4}\vert_{x = \sqrt{2 (\Omega/M-1)}}. 
\end{eqnarray}

\section{4-Channel Kondo effect}
\label{app:Kondo}
In this section, we explicitly derive the 4-channel Kondo physics from the effective $SU_2(2)$ theory and details on the noise magnetometry in the multichannel Kondo regime.

\subsection{Explicit mapping to 4-channel Kondo model}
The Kondo coupling, Eq.~\eqref{eq:Kondo}, involves $SU_2(2)$ currents which we represent as $J^a_{L/R}(x) = L^a_{bc}\chi_{L/R}^b(x) \chi_{L/R}^c(x)/2$ with $L^a_{bc} = -i \epsilon_{abc}$. We now transform Eq.~\eqref{eq:Kondo} into the standard form, in which only right moving currents couple to the spin. In this formulation, the analogy to the $M=3$ topological Kondo effect becomes apparent.

We first flip the direction of left currents by redefining Majorana fields $\chi_L(x) = \eta_R(-x)$. Then, we couple both rightmoving Majoranas into a Dirac fermion, i.e. $\Psi_R(x) = \chi_R(x) + i \eta_R(x) = \chi_R(x) +i \chi_L(-x)$. A calculation of the spin current yields

\begin{subequations}
\begin{align}
J^a(x) &= \frac{1}{4} \left \{{ \hat L^a(x)} + { \hat L^a(-x)} + {\tilde L^a(x)} - {\tilde L^a(-x) }\right \},\\
{ \hat L^a} &= { \Psi^\dagger L^a \Psi },\\
{\tilde L^a}&= {\frac{(\Psi^T L^a \Psi + \Psi^\dagger L^a \Psi^*)}{2}}.
\end{align}
\end{subequations}

In this formulation, the Kondo Hamiltonian becomes $H_{K} = J_K \sum_a J^a(0) \hat S^a$. We use the Majorana representation of the spin~\cite{Martin1959} $\hat S^a = - i \epsilon^{abc} \gamma_b \gamma_c$ where $\gamma_{a}, a \in \{1,2,3\}$ are Majorana fermions and $\hat S^a$ commutes with the overall parity $\gamma_0 \gamma_1 \gamma_2 \gamma_3$ (the physical subspace being the eigenstates with $\gamma_0 \gamma_1 \gamma_2 \gamma_3 =1$). Using $\epsilon^{abc} \epsilon^{ab'c'} = \delta^{bb'} \delta^{cc'} - (b \leftrightarrow c)$, Eq.~\eqref{eq:Kondo} becomes
\begin{equation}
H_{K} = 2 J_K \Psi^\dagger_a(0) \Psi_b(0) \gamma_b \gamma_c,
\end{equation}
which reproduces the basic model of the $M = 3$ topological Kondo effect, see e.g. Eq.~(1) of Ref.~\onlinecite{altland}.

To make connection to Gogolin and Fabrizio~\cite{FabrizioGogolin1994}, we switch to a basis in which $L^z$ is diagonal, using
\begin{equation}
U = \left (\begin{array}{ccc}
1/\sqrt{2} & 0 & -1/\sqrt{2} \\ 
i/\sqrt{2} & 0 & i/\sqrt{2} \\ 
0 & -1 & 0
\end{array}  \right)
\end{equation}
we obtain
\begin{eqnarray}
U^\dagger L^x U &=& \frac{1}{\sqrt{2}}\left (\begin{array}{ccc}
0 & 1 & 0 \\ 
1 & 0 & 1 \\ 
0 & 1 & 0
\end{array}  \right), \\ U^T L^x U &=& \frac{1}{\sqrt{2}}\left (\begin{array}{ccc}
0 & -1 & 0 \\ 
1 & 0 & 1 \\ 
0 & -1 & 0
\end{array}  \right),\\
U^\dagger L^y U &=& \frac{1}{\sqrt{2}}\left (\begin{array}{ccc}
0 & -i & 0 \\ 
i & 0 & -i \\ 
0 & i & 0
\end{array}  \right), \\ U^T L^y U &=& \frac{1}{\sqrt{2}}\left (\begin{array}{ccc}
0 & -i & 0 \\ 
i & 0 & -i \\ 
0 & i & 0
\end{array}  \right),\\
U^\dagger L^z U &=&\left (\begin{array}{ccc}
1 & 0 & 0 \\ 
0 & 0 & 0 \\ 
0 & 0 & -1
\end{array}  \right), \\
U^T L^z U&=& \left (\begin{array}{ccc}
0 & 0 & 1 \\ 
0 & 0 & 0 \\ 
-1 & 0 & 0
\end{array}  \right).
\end{eqnarray}
In this section we use the same bosonization convention as Ref.~\onlinecite{FabrizioGogolin1994}, $\psi^m(x) = [U^\dagger \Psi]^m(x) = \frac{1}{\sqrt{2\pi a}} e^{i \phi_m}$, where $m \in \{-1,0,1 \}$ is the spin quantum number. We thus obtain
\begin{eqnarray}
\hat L^x(x) &=&  \frac{\sqrt{2}}{\pi a} \cos \left(\sqrt{\frac{3}{2}} \phi_f \right) \cos \left(\sqrt{\frac{1}{2}} \phi_s \right),\\
\hat L^y(x) &=& - \frac{\sqrt{2}}{\pi a} \cos \left(\sqrt{\frac{3}{2}} \phi_f \right) \sin \left(\sqrt{\frac{1}{2}} \phi_s \right), \\
\hat L^z(x) &=&  \frac{1}{\sqrt{2} \pi} \partial_x \phi_s, \\
\tilde L^x(x) &=& \frac{\sqrt{2}}{\pi a} \cos \left({\frac{2}{\sqrt{3}} \phi - \frac{1}{\sqrt{6}}} \phi_f \right) \cos \left(\sqrt{\frac{1}{2}} \phi_s \right), \\
\tilde L^y(x) &=&- \frac{\sqrt{2}}{\pi a} \sin \left({\frac{2}{\sqrt{3}} \phi - \frac{1}{\sqrt{6}}} \phi_f \right) \cos \left(\sqrt{\frac{1}{2}} \phi_s \right),\\
\tilde L^z(x) &=&  \frac{1}{
\sqrt{2}\pi a} \cos\left (\frac{2}{\sqrt{3}} \phi + \frac{2}{\sqrt{6}} \phi_f \right).
\end{eqnarray}

Here, $\phi = (\phi_1 + \phi_0 + \phi_{-1})/\sqrt{3}$, $\phi_s = (\phi_1-\phi_{-1})/\sqrt{2}$, $\phi_f = (\phi_1-2 \phi_0 +\phi_{-1})/\sqrt{6}$, and all fields are evaluated at position $x$. At this point, we have mapped our model to the Fabrizio-Gogolin treatment of the four channel Kondo model and we can use their technique to calculate correlation functions. We remind the reader of their fixed point action in the rotated reference frame
\begin{eqnarray}
\tilde H &=& H_0 (\phi) + H_0 (\phi_s)+ H_0 (\phi_f) \notag \\
&&+ g_\perp S_x \cos(\sqrt{3/2} \phi_f) + \lambda \partial_x \phi_s S_z.
\end{eqnarray}

\subsection{Correlation Functions}
Far away from the impurity, all $\hat L$ and $\tilde L$ have trivial scaling dimension $d = 1$. Interestingly, to leading order in $\lambda \ll v$, $\hat L_z$ is unaffected by the impurity. For the purpose of noise magnetometry, we therefore disregard the contribution of local spin-currents of the wire. We specifically concentrate on the response of the spin-correlator (given by x-ray edge physics induced by spin flips $S_x \rightarrow - S_x$) and we consider the isotropic limit
\begin{equation}
\langle S^a(\tau) S^{b}(0) \rangle = \delta^{ab}  \underbrace{\mathcal C \left (\frac{\pi T/T_K}{\sin(\pi T \tau)} \right)^{2\Delta}}_{C^\tau(\tau)}.
\end{equation}
where in the case of the 4-channel Kondo effect $\Delta = 1/3$ and generically $\Delta = 2/(2+k)$~\cite{ParcolletSengupta1998}. 

For noise magnetometry, the retarded finite temperature response is needed and we Fourier transform of $C^\tau(\tau)$ ($\omega_n >0$).
\begin{subequations}
\begin{align}
C^\tau(z = i\omega_n) &= \mathcal C \int_0^\beta d\tau e^{z \tau}\left (\frac{\pi T /T_K}{\sin(\pi T \tau)} \right)^{2\Delta} \label{eq:TOCorrel}\\
&=  \mathcal C \int_0^\infty dt \Big[ e^{i z t} \left (\frac{\pi T /T_K}{-i\sinh(\pi T t)} \right)^{2\Delta} \notag \\
&- e^{i z (t + i \beta)} \left (\frac{\pi T /T_K}{-i\sinh(\pi T t + i \pi)} \right)^{2\Delta} \Big] \notag \\
&= \mathcal C  \sin(2\pi \Delta)\int_0^\infty dt e^{i z t} \left (\frac{\pi T /T_K}{\sinh(\pi T t)} \right)^{2\Delta}.
\end{align}
\end{subequations}
At this point we are ready to analytically continue the result, and obtain for $z \rightarrow \Omega + i\eta$
\begin{eqnarray}
C^R(z) &=&  -\mathcal C  \sin(2\pi \Delta) [-4^\Delta \Delta \Gamma(-2\Delta)] (\pi T/T_K)^{2\Delta-1} \notag \\
&&\times \frac{\Gamma(\Delta - i \frac{\Omega^+}{2\pi T})}{\Gamma(1-\Delta - i \frac{\Omega^+}{2\pi T})}.
\end{eqnarray} 

\subsection{Noise Magnetometry}

The energy relaxation rate of a single spin qubit in a fluctuating magnetic field is given by~\cite{RodriguezNievaDemler2018}
\begin{eqnarray}
\frac{1}{T_1} = \frac{(g \mu_B)^2}{2 \hbar^2} \coth\left (\frac{\hbar \Omega}{2 k_B T}\right) \text{ Im }[C_B^R(\Omega)].
\end{eqnarray}
Here, $\Omega$ is the frequency of the qubit and $C_B^R(\Omega)$ is the retarded Green's function of local magnetic field operators. The contribution from the impurity spin stems from dipole-dipole interactions
\begin{equation}
\v B(\v r, t) = \frac{\mu_0\mu_B}{4\pi} \frac{r^2 \v S(t) -3 \v r(\v S(t) \cdot \v r)}{r^5},
\end{equation}
and thus, 
\begin{eqnarray}
C^R_B(t) &=& - i \Theta(t) \langle [ B^a(\v r, t) , B^a(\v r, 0)] \rangle/3 \notag \\
&=& C^R(t)2 \left (\frac{\mu_0 \mu_B}{4\pi r^3}\right )^2.
\end{eqnarray}

We thus obtain a decay rate
\begin{eqnarray}
\frac{1}{T_1} &=& \mathcal C_\Delta \underbrace{\frac{\mu_0^2 g^2 \mu_B^4}{\hbar r^6} \frac{1}{k_B T_K}}_{\approx \frac{g^2 \times 8 Mhz}{r[nm]^6 T_K[K]}} \left (\frac{\pi T}{T_K} \right)^{2\Delta - 1} \notag \\
&&\times \underbrace{\coth \left (\frac{\Omega}{2\pi T} \right)\text{ Im}\left [ \frac{\Gamma(\Delta - i \frac{\Omega^+}{2\pi T})}{\Gamma(1-\Delta - i \frac{\Omega^+}{2\pi T})} \right]}_{\stackrel{\Omega \ll T}{\rightarrow} 1.14 }.
\end{eqnarray}
where $\mathcal C_\Delta \propto - 4^\Delta \Delta \sin(2\pi\Delta) \Gamma(-2\Delta)$ is a weakly $\Delta$ dependent constant of order unity. 
 

\section{Estimate of energy scales}
\label{app:EnergyScales}

We use Gaussian units in the presentation of electrostatic interaction energies, i.e. $4\pi \epsilon_0 =1$.

\subsubsection{Contact interaction}
We first estimate the bare energy scale of contact interaction. In the simplified case of a single 2D gate with screening length $l_{\rm TF}$ at distance $d$, the effective interaction in momentum space is~\cite{KoenigMirlin2013}
\begin{equation}
V (\v q) = \frac{2\pi e^2}{\epsilon} \frac{l_{\rm TF} + 2 d}{1+\vert \v q \vert l_{\rm TF}}.
\end{equation}
We estimate the typically transferred 2D momentum by $\vert \v q \vert \sim l^{-1}$, which is smaller than $d^{-1}$. Then we obtain in the experimentally realistic limit $l_{\rm TF} \ll d \lesssim l$
\begin{equation}
V_{\v x_1, \v x_2} = \frac{4\pi e^2 d}{\epsilon} \delta(\v x_1 - \v x_2)
\end{equation}
Using Eq.~\eqref{eq:InteractionMatrix} and the transverse width of wave functions $l$ this leads to an effective interaction constant $g = 4\pi e^2 d/(\epsilon l) \sim 27.5  d/(\epsilon l) \times 10^6 m/s$.

\subsubsection{Josephson coupling}
Next, we estimate the effective Josephson coupling $\lambda$. Single particle theories of small angle twisted bilayer graphene~\cite{mcdonald,FleischmannShallcross2019,TsimKoshino2020,DeBeuleRecher2020} suggest deflection coefficients which are larger than the corresponding transmission coefficients across an AA node of six domain walls. The physics behind this phenomenon was explored in Ref.~\onlinecite{QiaoNiu2014} and is related to the overlap of wavefunctions in adjacent wires. 
Without going into details, we here exploit the physical concepts of Ref.~\onlinecite{QiaoNiu2014} to estimate the single particle matrix elements between the wires which intersect at a given node. Then we use this to estimate the Josephson coupling.

At distance $r$ from the node, the hybridization (i.e. gap) between two adjacent wires is estimated by
$
t \sim m e^{- 2 r \sin(\pi/6) m/v}
$. Here, $m = m(r) = w_0 \tanh(r/l)$ is itself slowly position dependent, because the gap vanishes at the AA nodes (in the bulk of the wire, the gap is called $\delta_-$, see Appendix~\ref{app:MicroDerivation}). We remark that the size of the AA node is given by relaxation effects (estimated size $l$, up to 50 nm~\cite{polaritons}), while the decay length into the bulk can be much smaller $\xi_\infty = v/w_0 \sim 6$ nm. The optimal position for tunneling between adjacent wires ($r_{\rm opt}$) is given implicitly by the condition $\xi_\infty/l = \sinh ^2(r_{\rm opt}/l)+ \tanh (r_{\rm opt}/l)r_{\rm opt}/l \simeq 2r^2_{\rm opt}/l^2$. From this we obtain the maximal tunneling rate $t_{\rm opt} = w_0 \sqrt{\xi_\infty /(e 2l)} \sim 15$ meV to estimate the single particle hopping between adjacent wires.

Single particle hopping between adjacent wires is suppressed due to the dynamical mass gap $M \sim 20$ meV. We here extrapolate our treatment (valid for $t_{\rm opt} \ll M$) to the more realistic case $\lambda \lesssim M$, which should produce qualitatively correct results. The amplitude of pair hopping processes (i.e. Josephson coupling) can be estimated as $\vert\lambda\vert \sim t_{\rm opt}^2/M - e^2/(\epsilon l) \sim 1$ meV. Here, the first term stems from leading order perturbation theory in interwire hopping, while the second term accounts for the repulsive contribution of Coulomb interaction inside the AA node.

For the estimate of $T_c$ we further use $ v/L_\Theta \sim 3$ meV at $\theta = 0.06^\circ$ so that we obtain $T_c \sim 0.03$ meV (cf. Eq.~\eqref{eq:Tc}). This concludes the estimate of energy scales summarized in Table~\ref{tab:Scales} of the main text.

\bibliography{TBG_Network}
\end{document}